\documentclass[10pt,twocolumn,showkeys,footinbib]{revtex4-1}

\long\def\dontignore#1{#1}
\makeatletter

\long\def\ignoreflag{\@makeother\{\@makeother\}\xignore}
\long\def\xignore#1\dontignore#2{\catcode`\{\@ne
\catcode`\}\tw@
\afterassignment\xxdontignore\toks@\bgroup}
\long\def\xxdontignore{\the\toks@\ignoreflag}
\makeatother
\let\ignoreflag\relax

\usepackage{nicefrac}
\usepackage{tikz}
\usetikzlibrary{calc}
\usetikzlibrary{positioning}
\usetikzlibrary{decorations,positioning,shapes}
\usetikzlibrary{decorations.pathmorphing}

\usetikzlibrary{decorations.shapes}
\usetikzlibrary{decorations.markings}

\tikzstyle{vecArrow} = [thick, decoration={markings,mark=at position
   1 with {\arrow[semithick]{open triangle 60}}},
   double distance=1.4pt, shorten >= 5.5pt,
   preaction = {decorate},
   postaction = {draw,line width=1.4pt, white,shorten >= 4.5pt}]
\tikzstyle{innerWhite} = [semithick, white,line width=1.4pt, shorten >= 4.5pt]

\usepackage{natbib}
\usepackage{paralist}

\usepackage{amsmath}
\usepackage{amssymb}
\usepackage{import}

\usepackage{subfigure}

\usepackage{textcomp}
\usepackage[version=3]{mhchem}
\usepackage[utf8]{inputenc}

\usepackage{eurosym}

\usepackage{rotating}
\usepackage{multirow}

\definecolor{contour}{RGB}{16,16,127}
\definecolor{branchcut}{RGB}{255,63,0}

\newcommand{\manualcite}[1]{}

\newcommand{\di}{\mathrm{d}}

\newcommand{\e}{\,\mathrm{e}}
\newcommand{\erf}{\,\mathrm{erf}}

\newcommand{\order}[1]{\mathcal{O}(#1)}

\renewcommand{\imath}{{\mathrm i}}
\newcommand{\numberset}[1]{{\mathbb #1}}

\makeatletter\newcommand{\restoretitle}[1]{
\newcommand{\GNUPLOTspecial}{  \@sanitize\catcode`\%=14\relax\special}  {\GNUPLOTspecial{"
SDict begin [
  /Title (#1)
  /Subject (#1 Application)
  /Creator (LaTeX)
  /Author (Werner Koch)
  /CreationDate (Mon Mar  5 10:17:40 2012)
  /DOCINFO pdfmark
end
}}}
\makeatother

\usepackage[T1]{fontenc}

\DeclareUnicodeCharacter{2010}{-}

\newcommand{\mysection}[1]{\paragraph*{#1}}
\newcommand{\mysubsection}[1]{\paragraph*{#1}}

\providecommand{\keywords}[1]{\textbf{\textit{Index terms---}} #1}
\begin{document}
\ignoreflag

\date{\today}
\title{Systematic elimination of Stokes divergences emanating from complex phase space caustics}
\author{Werner Koch}
\affiliation{Weizmann Institute of Science, Rehovot, Israel}
\author{David J. Tannor}
\affiliation{Weizmann Institute of Science, Rehovot, Israel}

\begin{abstract}
Stokes phenomenon refers to the fact that the asymptotic expansion of complex functions can differ in different regions of the complex plane, and that beyond the so-called Stokes lines has an unphysical divergence.  An important special case is when the Stokes lines emanate from phase space caustics of a complex trajectory manifold. In this case, symmetry determines that to second order there is a double coverage of the space, one portion of which is unphysical. Building on the seminal but laconic findings of Adachi, we show that the deviation from second order can be used to rigorously determine the Stokes lines and therefore the region of the space that should be removed. The method has applications to wavepacket reconstruction from complex valued classical trajectories. With a rigorous method in hand for removing unphysical divergences, we demonstrate excellent wavepacket reconstruction for the Morse, Quartic, Coulomb and Eckart systems.
\end{abstract}
\maketitle

\mysection{Introduction}

Classical mechanics exhibits caustics, i.e. the focusing or coalescence of phase space trajectories when projected onto coordinate, momentum or any mixed subspace.
For real trajectories, this phenomenon is well understood \cite{delos_semiclassical_1986,littlejohn_semiclassical_1986,arnold1989mathematical} but in complex classical dynamics fundamental questions remain unanswered.
Extending the notion that caustics are stationary points of real trajectory manifolds, Adachi \cite{adachi_numerical_1989} and Rubin and Klauder \cite{rubin_comparative_1995} noted that caustics are second order saddle points of the complex trajectory manifold.
Unlike in purely real dynamics, an unphysical type of divergence arises in extended regions of the trajectory manifold that is connected to the caustics.
This type of divergence is called Stokes phenomenon, and the so-called Stokes lines, which emanate from the caustics in the complex space, define the boundary of the divergent contribution. \cite{stokes_discontinuity_1857,stokes_discontinuity_1902}.

Stokes divergences are a major obstacle in semiclassical wavepacket propagation methods that employ complex trajectories. Significant progress has been made towards eliminating such divergences in discrete time systems based on Adachi's work on the Principle of Exponential Dominance (PED) \cite{adachi_numerical_1989,shudo_complex_1995,shudo_stokes_1996,shudo_toward_2016}.
However the PED requires comparison of trajectory pairs which can only be found numerically via a root search in the complex trajectory manifold.
In continuous time dynamics such a root search is quite challenging and has been successful so far only for relatively short times \cite{rubin_comparative_1995,parisio_regular_2005,aguiar_semiclassical_2005,goldfarb_complex_2008,petersen_wave_2015}.
Long time dynamics where the trajectory manifold is plagued by a multitude of caustics, remains elusive.

In this Communication, we provide a practical and rigorous procedure for removing Stokes divergences in continuous time complex trajectory manifolds. Building on the seminal but laconic work of Adachi, we perform a local expansion around the caustics. We show that the deviation from symmetry beyond second order can be used to rigorously determine the position of the Stokes lines and therefore the divergent region of the space that should be removed. Combining the method with the final value coherent state propagator (FINCO) approach \cite{zamstein_communication:_2014} we calculate wavepacket dynamics for the Morse, Quartic, Coulomb and Eckart systems in excellent agreement with the quantum results.

\mysection{Double-valuedness in the neighborhood of caustics}
\label{sec:caustics}

Consider a complex valued Lagrangian manifold of initial conditions, $p_0(\nu),q_0(\nu)\in\numberset{C}$ at time $t_0$, where $\nu$ is the manifold label.
An important example is the (possibly non-linear) manifold determined through a complex function $S(x)$ by $q_0(\nu)=\nu,p_0(\nu)=\frac{\partial S(x)}{\partial x}\vert_{x=\nu}$.
This manifold is propagated using Hamilton's equations of motion leading to the manifold $p_t(\nu),q_t(\nu)\in\numberset{C}$ at time $t$. Choosing parameters $\alpha,\beta\in\numberset{C}$, we define a map $\xi(\nu)$:
\begin{align}
\label{eq:projection}
  \xi(\nu)=\alpha q_t(\nu) + \frac{1}{\hbar}\beta p_t(\nu)\,.
\end{align}
For example $\alpha=1,\beta=0$ defines a map onto coordinate space at time $t$.

The map Eq.~\eqref{eq:projection} has a caustic at $\nu^{\ast}$ if
\begin{align}
\label{eq:caustic}
   \xi^{(1)}(\nu^{\ast})\equiv\left.\frac{\di \xi(\nu)}{\di\nu}\right|_{\nu=\nu^{\ast}}=0\,.
\end{align}
Such locations always exist independent of $\alpha$ and $\beta$ if the manifold $p_t(\nu),q_t(\nu)$ is a nonlinear function of $\nu$.
Higher order caustics may also exist but will not be considered in this Communication.

Since $\xi^{(1)}(\nu^{\ast})=0$, the function $\xi(\nu)$ is locally quadratic at the caustic, $\xi(\nu)=\xi(\nu^{\ast})+\frac{1}{2}\xi^{(2)}(\nu^{\ast})(\nu-\nu^{\ast})^2+\order{(\nu-\nu^{\ast})^3}$.
This implies the existence of values $\nu_1,\nu_2$ in the vicinity of $\nu^{\ast}$ for which
$\xi(\nu_1)=\xi(\nu_2)$,
i.e. $\xi(\nu)$ is a double cover near $\nu^{\ast}$ \cite{adachi_numerical_1989,rubin_comparative_1995}.
Conversely, the inverse function $\nu(\xi)$ is double valued near $\xi^{\ast}=\xi(\nu^{\ast})$.

Consider a function $L(\xi)$ defined through the trajectory manifold in the form
\begin{align}
\label{eq:SCSP}
  L(\xi)=\sum_{j}\phi_j(\xi(\nu_j))\e^{\sigma_j(\xi(\nu_j))}\,,
\end{align}
where $\phi_j(\nu)$ and $\sigma_j(\nu)$ are complex valued and the sum is over all trajectories in the manifold for which $\xi(\nu_j)=\xi$.
Since the trajectory manifold $\xi(\nu)$ is a double cover in the vicinity of a caustic, the sum in Eq.~\eqref{eq:SCSP} for a given $\xi$ contains two terms
 corresponding to the manifold labels $\nu_1$ and $\nu_2$.
Sums of the form Eq.~\eqref{eq:SCSP}  occur in semiclassical expressions for the  quantum mechanical propagator where the sum is over root trajectories, \cite{klauder_recent_1987},
 as well as in semiclassical methods for calculating wavepacket evolution \cite{huber_generalized_1987,aguiar_semiclassical_2005,de_aguiar_initial_2010,zamstein_communication:_2014}.

\mysubsection{Stokes divergences}
\label{sec:stokes}

Since each of the two classical trajectories represents a distinct root trajectory, the dependence of $\{\sigma_1,\phi_1\}$ and $\{\sigma_2,\phi_2\}$ on $\xi$ and $\nu$ may be very different globally.
Specifically, in regions of the complex $\xi$ plane where the saddle point approximation used to obtain Eq.~\eqref{eq:SCSP} is invalid, unphysical divergences of $\phi(\xi)\e^{\sigma(\xi)}$ may appear, known as Stokes phenomenon.
This change of the validity of the approximation \cite{stokes_discontinuity_1857,stokes_discontinuity_1902} can be characterized by the Stokes variable, defined to be the difference of the exponents of the two terms
\begin{align}
\label{eq:stokes_variable}
  F(\xi)=\sigma_1(\xi)-\sigma_2(\xi)\,.
\end{align}
Starting from a central point $\xi^{\ast}$ where $F(\xi^{\ast})=0$, the surrounding space is divided into sectors by the Stokes lines, the locus of points where $\Im F(\xi)=0$, and the anti-Stokes lines, the locus of points where $\Re F(\xi)=0$ \footnote{In the literature, these definitions also occur reversed. We use the definition of Ref.~\cite{berry_uniform_1989}}.
By definition, on the anti-Stokes lines $\Re \sigma_1=\Re \sigma_2$ and thus the exponential part of the two terms in Eq.~\eqref{eq:SCSP} is of equal magnitude $|\e^{\sigma_1}|=|\e^{\sigma_2}|$.
On the Stokes lines, on the other hand, $\Im \sigma_1=\Im \sigma_2$ and the exponential parts $\e^{\sigma_1}$ and $\e^{\sigma_2}$ are of equal phase but differ in magnitude.

If one of the contributions to Eq.~\eqref{eq:SCSP} diverges unphysically in a certain region of the complex $\xi$ plane, this contribution needs to be excluded and the sum in Eq.~\eqref{eq:SCSP} is reduced to the other term.
The most straightforward way to eliminate the Stokes divergence is simply to discard trajectories with $\Re \sigma(\xi)>0$, corresponding to the invalid complex integration contour \cite{adachi_numerical_1989,rubin_comparative_1995}.
However, simply removing trajectories with $\Re \sigma(\xi)>0$ leads to inaccurate numerical results,
due to the rapidly fluctuating phase $\e^{\imath\Im\sigma(\xi)}$ at the boundary $\Re \sigma(\xi)=0$.
A number of criteria have been proposed to remove divergences based on the value of $\sigma(\xi)$ or components thereof \cite{aguiar_semiclassical_2005,parisio_regular_2005,zamstein_communication:_2014,koch_wavepacket_2017},
but these are empirical at best and do not lead to satisfactory results for long time propagation.

Adachi ~\cite{adachi_numerical_1989} observed that a divergent region of the complex $\xi$ plane contains a Stokes line where $\Re \sigma_2(\xi)$ is maximal and $\Re \sigma_1(\xi)$ is minimal.
He noted that removing all trajectories up to the neighboring Stokes lines on either side, where $\Re \sigma_2(\xi)$ is minimal and $\Re \sigma_1(\xi)$ is maximal,
results in minimum discontinuity because the removed contribution $\e^{\sigma_2(\xi)}$ is masked by the exponentially larger term $\e^{\sigma_1(\xi)}$ at the Stokes line.
The discontinuity can be removed entirely by using Berry's smoothing factor  $S(F(\xi))=\erf \left\{\frac{\Im F(\xi)}{\sqrt{2\Re F(\xi)}}\right\}$,
 derived from asymptotic analysis \cite{berry_uniform_1989}:
\begin{align}
\label{eq:smooth_xi}
\tilde{L}(\xi)=\phi_1(\xi)\e^{\sigma_1(\xi)} + S(F(\xi)) \phi_2(\xi)\e^{\sigma_2(\xi)}
\end{align}
where $\erf(\tau){=}\frac{1}{\sqrt{\pi}}\int_{-\infty}^\tau\di{s}\e^{-s^2}$ is the error function.
As illustrated below in Fig.~\ref{fig:stokesLines}, in Berry's procedure trajectories are removed completely only from the divergent Stokes line to the adjacent \emph{anti} Stokes lines followed by a continuous transition from excluded to included in the neighboring anti Stokes sector.

Below we combine Adachi's removal of the diverging contribution with Berry's smoothing factor. But the procedure requires knowing the location of the Stokes lines.
A brute force approach is to use a root search to determine for each trajectory $\nu_1,\xi(\nu_1)$ its conjugate trajectory $\nu_2$ fulfilling $\xi(\nu_1)=\xi(\nu_2)$, from which one can locate the Stokes lines. This procedure was apparently applied in Refs.~\cite{huber_generalized_1988,shudo_complex_1995,shudo_stokes_1996,goldfarb_complex_2008,petersen_wave_2015}, but becomes cumbersome for long propagation times as the number of caustics proliferates ~\cite{shudo_stokes_1996}.
The main achievement of this paper is to provide a practical procedure to locate the boundary between included and excluded contributions.
This is the subject of the next section.

\mysection{Systematic elimination of the Stokes divergence by identifying the double cover}
\label{sec:systematic elimination}

A much more efficient approach to eliminate the Stokes divergence exploits an analytic expansion of the Stokes variable $F$ in terms of $\nu$ and $\xi$.
This expansion provides $F(\xi)$ without the need to search for conjugate trajectories.

The condition for the double cover $\xi(\nu_1)=\xi(\nu_2)$  implicitly defines a relationship
  $\nu_2(\nu_1)$
for the conjugate manifold label $\nu_2$. Combining $\nu_2(\nu_1)$ with the map $\xi(\nu)$, we transform the Stokes variable from $\xi$ space to $\nu$ space
\begin{align}
\label{eq:stokes_variable_nu}
  F(\nu_1)=\sigma(\nu_1)-\sigma(\nu_2(\nu_1))\,.
\end{align}
Whereas previously it was necessary to distinguish between $\sigma_1(\xi)$ and $\sigma_2(\xi)$ since $\xi(\nu)$ is a double cover, as a function of $\nu$ $\sigma(\nu)$ is single-valued.

Since $\sigma(\nu)$ and $\xi(\nu)$ are single-valued functions, we can use their Taylor expansions at the caustic $\nu^{\ast}$
to construct an approximation to Eq.~\eqref{eq:stokes_variable_nu} and determine  Stokes lines and anti Stokes lines from that.
Without loss of generality, in the following we will assume $\nu^{\ast}=0$ and $\xi(\nu^{\ast})=0$.

\begin{figure}[htp]
\begin{center}
  \includegraphics{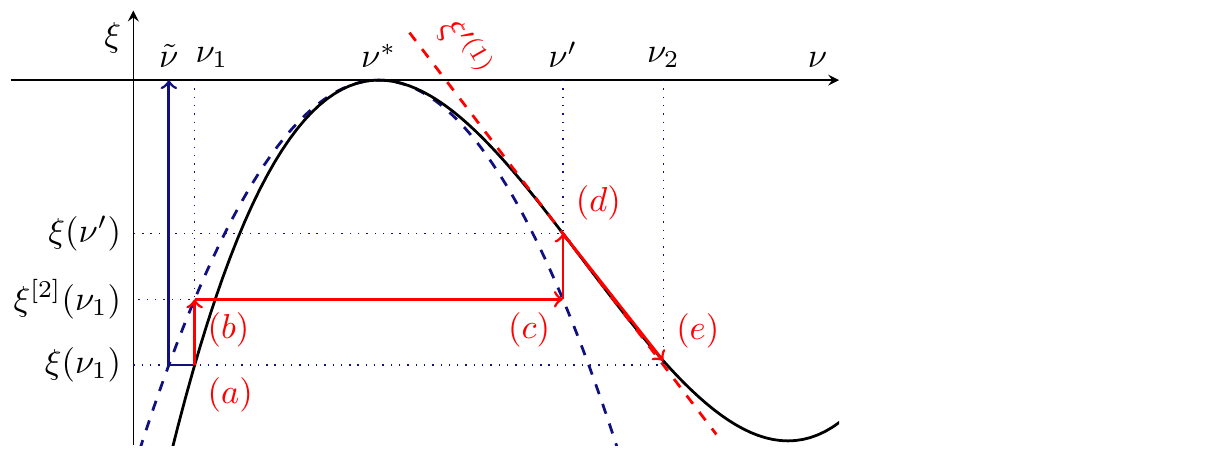}
\end{center}
  \caption[]{Locating the conjugate point $\nu_2(\nu_1)$ for a manifold label $\nu_1$ such that $\xi(\nu_1)=\xi(\nu_2)$ in the local double cover $\xi(\nu)$ (black, solid):\\
(a) Start at the initial point $\nu_1,\xi(\nu_1)$\\
(b) Move to second order expansion  $\xi^{[2]}(\nu_1)$ (blue, dashed)\\
(c) Reflect at $\nu^{\ast}$ to get approximate $\nu',\xi^{[2]}(\nu')$\\
(d) Move to full expansion $\nu',\xi(\nu')$\\
(e) Shift to $\nu'+[\xi'^{(1)}]^{-1}\Delta\xi+\order{\Delta\xi^2}$ using derivative $\xi'^{(1)}(\nu')$ and offset $\Delta\xi=\xi(\nu_1)-\xi(\nu')$
(red, dashed)\\
Note that in reality, both $\nu$ and $\xi$ are complex valued.
}
\label{fig:causticMirror}
\end{figure}

The first step is to obtain an explicit form for the conjugate point $\nu_2(\nu_1)$.
Since $\xi(\nu)$ is a double cover in the vicinity of the caustic, a straight-forward inversion is impossible.
Instead, we note that the second order Taylor expansion $\xi^{[2]}(\nu)$ is symmetric.
Thus we define the estimate conjugate label
$\nu'  =  -\nu_{1}$,
and obtain corrections to that estimate using a local inversion of $\xi(\nu)$ at $\nu'$.
The approach is illustrated in Fig.~\ref{fig:causticMirror}.

The conjugate point $\nu_2$ follows from the Taylor expansion of $\nu(\xi)$ (the inverse of $\xi(\nu)$) at $\xi'\equiv\xi(\nu')$
\begin{align}
\label{eq:taylor_nu_2}
  \nu_2(\nu_1)&=\nu'+\sum_{i=1}^{\infty}\frac{\nu^{(i)}(\xi')}{i!}\left(\Delta\xi(\nu_1)\right)^i\,,
\end{align}
where the deviation $\Delta\xi(\nu_1)$ is defined as:
\begin{align}
\label{eq:deltaXi}
\Delta\xi(\nu_1) \equiv & \xi\left(\nu_{1}\right)-\xi\left(\nu'\right)= \frac{\xi^{(3)}(\nu^{\ast})}{3}\nu_{1}^{3}+\order{\nu_{1}^{5}}\,.
\end{align}

The derivatives of the inverse $  \nu^{(i)}(\xi')$ are obtained from the derivatives of the Taylor series evaluated at $\nu'$
\begin{align}
\xi'^{(i)}
  \equiv &\left.\frac{\di^i \xi(\nu)}{\di \nu^i}\right|_{\nu=\nu'} =\sum_{j=i}^{\infty}\frac{\xi^{(j)}(\nu^{\ast})}{(j-i)!}(\nu'-\nu^{\ast})^{j-i}
\end{align}
by applying the chain rule to the identity $\nu=\xi^{-1}(\xi(\nu))$.
As an example, the first two of these evaluate to
\begin{subequations}
\label{eq:inverse_derivs}
\begin{align}
  \nu^{(1)}(\xi')&=[-\xi^{(2)}(\nu^{\ast})\nu_{1}+\order{\nu_{1}^{2}}]^{-1}\\
\nu^{(2)}(\xi')&=-\frac{-\xi^{(2)}(\nu^{\ast})+\order{\nu_{1}}}
{[-\xi^{(2)}(\nu^{\ast})\nu_{1}+\order{\nu_{1}^{2}}]^3}\,.
\end{align}
\end{subequations}
Note that the derivatives of the inverse involve the inverses of the derivatives.
Inserting Eqs.~\eqref{eq:inverse_derivs} and~\eqref{eq:deltaXi} into Eq.~\eqref{eq:taylor_nu_2} and re-expanding the quotients in terms of $\nu_1$ yields
\begin{align}
\label{eq:nu_2_expansion}
  \nu_2(\nu_1)=-\nu_1+\varrho\nu_1^2-\varrho^2\nu_1^3+\order{\nu_1^4}
\end{align}
with $\varrho=\frac{-\xi^{(3)}(\nu^{\ast})}{3\xi^{(2)}(\nu^{\ast})}$.

The second step is to insert Eq.~\eqref{eq:nu_2_expansion} and the Taylor series of $\sigma(\nu)$ into Eq.~\eqref{eq:stokes_variable_nu} and we obtain the final result, the phase space caustic Stokes expansion (PCSE) in terms of the manifold label
\begin{align}
\label{eq:taylor_stokes_variable}
  F(\nu_1)=F^{(3)}\nu_1^3+\order{\nu_1^4}\,,
\end{align}
where $F^{(3)}=\frac{1}{3}\sigma^{(3)}(\nu^{\ast})+\sigma^{(2)}(\nu^{\ast})\varrho$.

Before discussing the use of $F(\nu_1)$ for locating Stokes lines,
we describe some of its properties.
\begin{asparaenum}[a)]
\item
Assuming that $\sigma(\nu)$ has no explicit $\nu$ dependence, its first derivative $\sigma^{(1)}(\nu^{\ast})=\frac{\partial \sigma(\xi)}{\partial \xi(\nu)}\xi^{(1)}(\nu^{\ast})$ vanishes because by definition Eq.~\eqref{eq:caustic}, at the caustics  $\xi^{(1)}(\nu^{\ast})=0$ vanishes.
Moreover, due to the symmetry of $\sigma(\nu_1)$ and $\xi(\nu_1)$, Eq.~\eqref{eq:stokes_variable_nu} vanishes up to second order.
\item
The dominant term in Eq.~\eqref{eq:taylor_stokes_variable} is of third order.
Hence there are six lines emanating from $\nu^{\ast}$ where $\Im F=0$.
In $\xi$ space, this leads to three Stokes lines enclosing angles $\frac{2\pi}{3}$.
This term appears in a comment in Ref.~\cite{adachi_numerical_1989} without derivation but correctly identifying the three Stokes lines.
\item
The third order coefficient $F^{(3)}$ critically depends on the correction $\varrho\nu_1^2$ to the estimate $\nu'=-\nu_1$ in Eq.~\eqref{eq:nu_2_expansion}.
Without it (equivalent to setting $\varrho=0$), the direction of the Stokes lines predicted by Eq.~\eqref{eq:taylor_stokes_variable} will be incorrect.
\item
Due to symmetry, third order information at the caustic, i.e. $\xi^{(i)}(\nu^{\ast}),\sigma^{(i)}(\nu^{\ast}),i\in\{2,3\}$ is sufficient for a fourth order accurate approximation.
The fourth order coefficient is $F^{(4)}=-\frac{3\varrho}{2}F^{(3)}$.

\item
The expansion \eqref{eq:taylor_stokes_variable} in terms of the manifold label $\nu_1$ was derived using the expansion of $\xi(\nu)$ at the caustic.
It turns out that for long propagation times and trajectories far from the caustic,
it is more accurate to expand in terms of $\tilde{\nu}(\nu_1)=\pm\left(\frac{\xi(\nu_1)}{\xi^{(2)}(\nu^{\ast})}\right)^{\frac{1}{2}}$
\begin{align}
\label{eq:taylor_F_xi}
\tilde{F}(\nu_1)=F^{(3)}\tilde{\nu}(\nu_1)^{3}+\order{\tilde{\nu}(\nu_1)^5}\,,
\end{align}
with the sign of $\tilde{\nu}$ chosen such that $|\tilde{\nu}(\nu_1)-\nu_1|$ is minimal.
The geometric interpretation of $\tilde{\nu}$ is given in Fig.~\ref{fig:causticMirror}.
\end{asparaenum}

With Eq.~\eqref{eq:taylor_F_xi} in hand, the following procedure removes Stokes divergences from the trajectory manifold without the need to find conjugate trajectories (dropping the subscript of $\nu_1$).
\begin{compactenum}[a)]
\item
\label{it:manifold}
Construct the initial manifold $p_0(\nu),q_0(\nu)$.
\item
Propagate to $p_t(\nu),q_t(\nu)$ and compute the final phase space map $\xi(\nu)$, the prefactors $\phi(\nu)$ and exponents $\sigma(\nu)$,
 and the stability matrix elements $M_{ab}=\frac{\partial a_t(\nu)}{\partial b_0(\nu)}$ with $a,b\in\{p,q\}$.
\item
Locate the caustics in the manifold as the roots of
\begin{align}
  \label{eq:DxiDnu}
\hspace{-.3em}\frac{\di \xi(\nu)}{\di \nu}
=\left(\alpha M_{qp}{+}\beta M_{pp}\right)\frac{\partial p_0}{\partial \nu}{+}\left(\alpha M_{qq}{+}\beta M_{pq}\right) \frac{\partial q_0}{\partial \nu}\,.
\end{align}
\item
\label{it:PCSE}
For each caustic $\nu^{\ast}$:
\begin{asparaenum}[i]
\item
Compute $\xi^{(i)}(\nu^{\ast}),\sigma^{(i)}(\nu^{\ast}),i\in\{2,3\}$ via finite differencing with trajectories near $\nu^{\ast}$.
\item
Split the manifold into six sectors along the anti Stokes lines $\Re\tilde{F}(\xi(\nu))=0\,.$ (See Fig.~\ref{fig:stokesLines}.)
\item
Remove the sector that contains the Stokes line $\Im\tilde{F}(\xi(\nu))=0$ along which $\Re\sigma(\nu)>0$ diverges.
\item
Multiply all trajectory contributions $\phi(\nu)\e^{\sigma(\nu)}$ in the two adjacent sectors by the Berry factor $S(\tilde{F}(\xi(\nu)))$ according to Eq.~\eqref{eq:smooth_xi}.
\end{asparaenum}
\end{compactenum}

\vspace{1EM}
\mysubsection{Application to wavepacket reconstruction}

We illustrate this procedure by using it for wavepacket reconstruction in the context of the final value coherent state propagator method (FINCO) \cite{zamstein_communication:_2014}.
The initial manifold is derived from the analytic continuation of a wavefunction $\psi_0(x)$ according to $q_0(\nu)=\nu$ and $p_0(\nu)=-\imath\hbar\left.\frac{\partial}{\partial x}\ln\left\{\psi_0(x)\right\}\right|_{x=q_0(\nu)}$.
The parameters of the map Eq.~\eqref{eq:projection} are  $\alpha=2\gamma$ and $\beta=-\imath$ with $\gamma\in\numberset{R}^{+}$ an arbitrary real, positive parameter.
The exponents have the form
\begin{align}
\label{eq:sigma}
\sigma(\nu)&=\frac{\imath}{\hbar} S_{t}(\nu)+\frac{1}{4\gamma\hbar^2}p_t^2(\nu)-\frac{1}{4\gamma}\bigl(\Im \xi(\nu)\bigr)^2
\end{align}
where $S_{t}(\nu)$ is the classical action integrated along the trajectory.
The last term controls normalization and is \emph{not} an analytic function of $\nu$ (see e.g. Ref.~\cite{baranger_semiclassical_2001}).
However, since we need only the difference Eq.~\ref{eq:stokes_variable} with $\xi(\nu_1)=\xi(\nu_2)$, we may modify $\sigma(\nu)$ by any purely $\xi$ dependent quantity without changing $F(\nu)$.
Thus we use the analytic function $\sigma_{A}(\nu)=\frac{\imath}{\hbar} S_{t}(\nu)+\frac{1}{4\gamma\hbar^2}p^2_t(\nu)$ with derivative $\sigma_{A}^{(1)}(\nu)=\frac{\imath}{2\gamma\hbar}p_t(\nu)\xi^{(1)}(\nu)$ in step~\ref{it:PCSE}) for determining $\tilde{F}(\xi(\nu))$.

We choose a Gaussian initial wavepacket \begin{align}
  \label{eq:gaussian}
\psi_0(x)=g_{\gamma_0}(x,\chi) = \left(\tfrac{2\gamma_0}{\pi}\right)^{\frac{1}{4}}\e^{ -\gamma_0\left( x-\tfrac{\chi^{\ast}}{2\gamma_0}\right)^{2}
-\tfrac{\left(\Im\chi\right)^{2}}{4\gamma_0}}
\end{align}
where ${}^{\ast}$ denotes the complex conjugate and $\chi=-2\imath$ and $\gamma_0=\frac{1}{2}$.
Trajectories are propagated until time $t=0.5$ in a Quartic potential (see Table~\ref{tab:parameters}). As shown in Fig.~\ref{fig:stokesLines}, the Stokes lines obtained for $\gamma=\frac{1}{2}$ through the procedure described above agree very well with those obtained by root search in the trajectory manifold.
\begin{figure}[htp]
\begin{center}
  \includegraphics{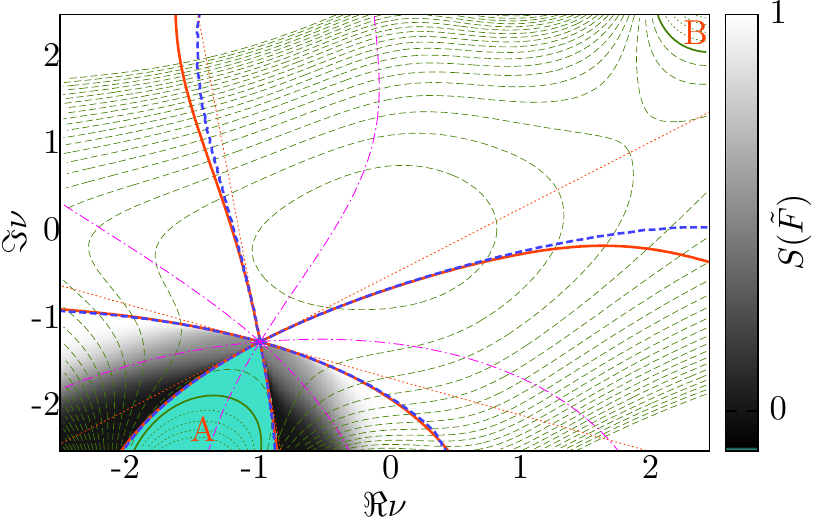}
  \end{center}
\caption{
\label{fig:stokesLines}
Anti Stokes lines in a trajectory manifold for the Quartic potential
with a caustic at $\nu^{\ast}=-0.96-1.24\imath$ (see Table~\ref{tab:parameters} for details).
The root search result (blue, dashed) is well approximated by the expansion Eq.~\eqref{eq:taylor_F_xi} (red, solid) and approximated locally by the third order term of Eq.~\eqref{eq:taylor_stokes_variable} (red, thin dotted).
Each anti Stokes sector is bisected by a Stokes line (purple, dash-dotted).
Two regions with $\Re\sigma(\nu)>0$ are marked with letters {\color{branchcut}A,B}, bounded by thick green lines.
The anti Stokes sector containing region {\color{branchcut}A} is removed (turquoise, below the caustic).
The shade in the adjacent sectors shows Berry's Stokes multiplier $S(\tilde{F})$.
Note that region {\color{branchcut}B} is associated with a second caustic at $\nu^{\ast}=1.95+1.13\imath$.
}
\end{figure}

 In FINCO, a time evolved wavepacket is computed using a Gaussian basis $g_\gamma(x,\xi)$ with $\gamma=\frac{1}{2}$ (analogous to Eq.~\eqref{eq:gaussian}) with coefficients given by Eq.~\eqref{eq:SCSP}
\begin{align}
\label{eq:SC_xi}
  \psi_t(x)=-\frac{1}{4\pi\gamma}\int\di\xi g_\gamma(x,\xi)L(\xi)\,.
\end{align}
 The sum over trajectories in $L(\xi)$ is accounted for by transforming to $\nu$ space
\begin{align}
\label{eq:SC_nu}
\hspace{-0.25em}\psi_t(x)=\frac{-1}{4\pi\gamma}\int\di\nu |J(\nu)| g_\gamma(x,\xi(\nu))\phi(\nu)\e^{\sigma(\nu)}\,,
\end{align}
with the Jacobian
$|J(\nu)|=\left|\xi^{(1)}(\nu)\right|^2$
 and prefactor
$\phi(\nu)=\left(8\gamma\pi\right)^{\frac{1}{4}}\left[\xi^{(1)}(\nu)\right]^{-\frac{1}{2}}$ where $\xi^{(1)}(\nu)$ is given by Eq.~\eqref{eq:DxiDnu}.
The exponent $\sigma(\nu)$ is defined in Eq.~\eqref{eq:sigma}.
For more details see Refs.~\cite{zamstein_communication:_2014,koch_wavepacket_2017}.

\begin{figure}[htp]
  \centering
\resizebox{\columnwidth}{!}{
    \includegraphics{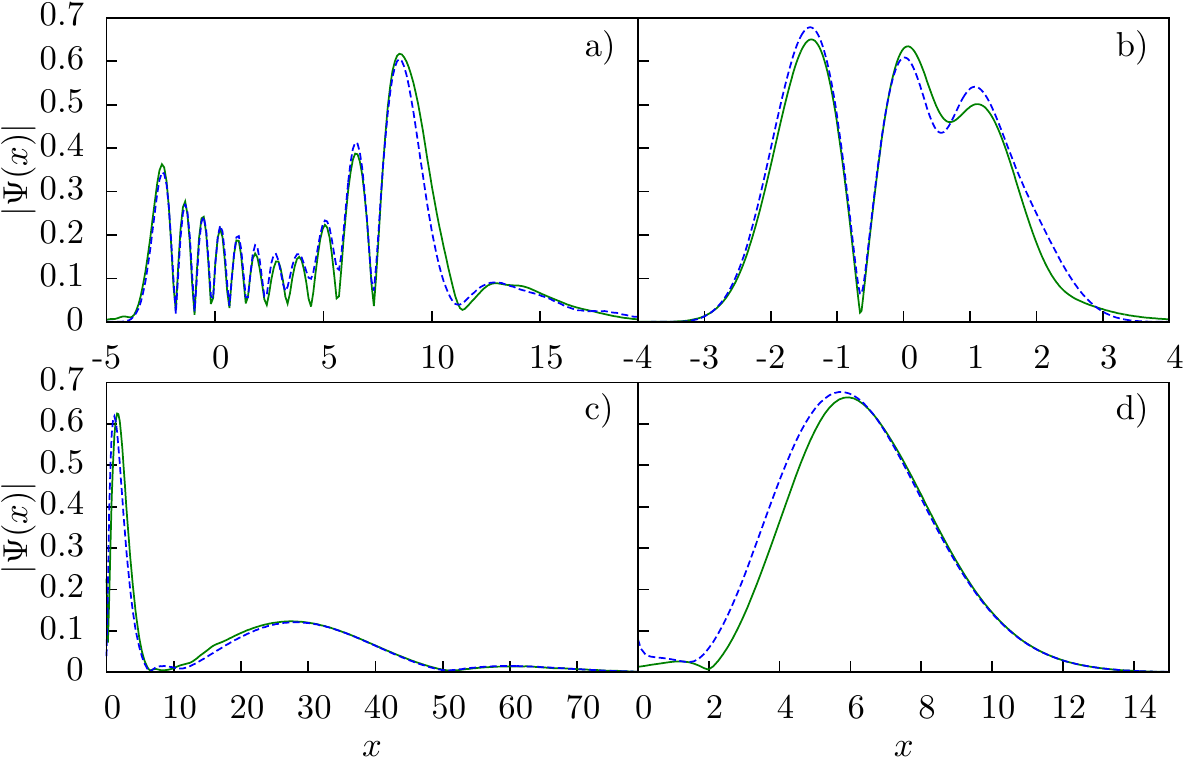}
   }
  \caption{Trajectory based wavefunctions (solid, green) and numerical quantum results (dashed, blue) at time $t=3\,T_{\rm cl}$ in a) Morse, b) Quartic and c) Coulomb potential.
For the Eckart barrier d), only the transmitted part of the wavepacket is shown enlarged by a factor of $10$ at $t\approx1.5\,T_{\rm cl}$.
For parameters see Table~\ref{tab:parameters}.
}
  \label{fig:wavepackets}
\end{figure}

Trajectory based wavepacket reconstructions are compared with quantum results in Fig.~\ref{fig:wavepackets}.
The initial wavepackets Eq.~\eqref{eq:gaussian} with width $\gamma_0=\frac{1}{2}$ are propagated for three classical periods of oscillation in the Morse, Quartic and Coulomb potentials and for about $1.5$ times the classical turning point time in the Eckart potential.
Potential functions and system parameters are given in Table ~\ref{tab:parameters}.
The quality of the reproduction is semi-quantitative throughout.

Note that for the Morse, Coulomb and Eckart systems, imaginary time contour propagation is required to obtain all relevant contributions \cite{petersen_wave_2015,koch_wavepacket_2017,koch_multivalued_2018}.
A separate publication will give more details of the implementation and the phase space geometry of the caustics in the four prototypical potentials.

In summary, the present paper provides a practical and rigorous way to remove Stokes divergences emanating from caustics in complex trajectory manifolds. Besides its inherent conceptual interest it opens the door to accurate, longtime wavepacket propagation using complex valued classical trajectories.

This work was supported by the Israel Science Foundation (1094/16) and the German-Israeli Foundation for Scientific Research and Development (GIF).

\begin{table}[htp]
  \caption{Potentials $V(x)$, phase space centers $\chi$, classical reference times $T_{\rm cl}$ and masses $m$ for wavepacket reconstructions shown in Fig.~\ref{fig:wavepackets}. The references given are the sources of the parameters and all values are in atomic units.}
  \label{tab:parameters}
  \begin{tabular}[c]{lc|cccc}
Potential & $V(x)$ & $\chi$ &  $T_{\rm cl}$ & $m$ & Ref.\\
\hline
Morse & $10.25\left(1-\mathrm{e}^{-0.2209 x}\right)^{2}$ & 9.342 & $12.88$ & 1 & \cite{huber_generalized_1987}\\
Quartic & $\frac{1}{2}x^{2}+\frac{1}{10}x^{4}$ &$-2\imath$ & 4.72 & 1 & \cite{de_aguiar_initial_2010}\\
Coulomb & $-\frac{1}{x}$ & 2 & $2\pi$ & 1 & n.a.\\
Eckart & $0.01562\cosh(x/0.734)^{-2}$ & $-8+4\imath$ & 2113 & 1060 & \cite{petersen_wave_2015}
  \end{tabular}
\end{table}

\restoretitle{Systematic elimination of Stokes divergences emanating from complex phase space caustics}
\bibliography{manuscript}
\bibliographystyle{apsrev}
\dontignore{\end{document}}